# Arbitrary spin-controlled directional coupling of surface plasmon polaritons

**Hailong Zhou,[1,2] Jinran Qie,[1,2] Jianji Dong[1,*] and Xinliang Zhang[1,*]**

[1]Wuhan National Laboratory for Optoelectronics, School of Optoelectronic Science and Engineering, Huazhong University of Science and Technology, Wuhan 430074, P. R. China
[2]These authors contributed equally to this work
*Correspondence to: J. Dong (e-mail: jjdong@mail.hust.edu.cn)
*Correspondence to:X. Zhang (e-mail: xlzhang@mail.hust.edu.cn).

Abstract: Controlling the directionality of surface plasmon polaritons (SPPs) has been widely studied, while the direction of SPPs was always switched by orthogonal polarizations in the reported methods. Here, we propose a scheme to control the directionality of SPPs for arbitrary input of spin polarizations. Specifically, the device can split two quite adjacent polarization components to two opposite directions. The versatility of the presented design scheme can provide opportunities for polarization sensing, polarization splitting and polarization-controlled plasmonic devices.

## 1. Introduction

Surface plasmon polaritons (SPPs) are electromagnetic excitations constituted by a light field coupled to a collective electron oscillation, and they propagate along the interface between a metal and a dielectric [1]. They have attracted lots of research interests because of their intriguing properties such as field localization and enhancement. Nowadays, the SPPs have been widely used in a variety of interesting applications, such as plasmon focusing [1-9], plasmonic complex-field generation [10-12], directional coupling [13-18].

One of these important branches of researches about SPPs is the directional coupling controlled by input polarizations. Various polarization-controlled plasmonic directional launching methods were reported. For example, using the meta-slit structures composed of arrays of subwavelength-scale nanoslit segments, the SPPs could be alternatively launched into the two opposite directions by switching the handedness of the incident circularly polarized light [13, 14, 19]. Subsequently, various derivative applications were exploited, such as plasmonic Lens [9, 20], polarimeter [21, 22] ,plasmonic complex-field shaping [12, 16, 23], reflectionless directional converters between freely propagating visible photons and SPPs [24], and plasmonic holography [25]. Other structures of polarization-controlled directional coupling of SPPs were also covered, like utilizing U-shaped subwavelength plasmonic waveguides [26], single nanoslit [27], ∆-nanoantennas [17], split-ring-shaped slit resonators[15] and arrays of gap SPP resonators [28]. These methods can extend the applicationsin polarization-sensitive SPPs excitation and polarization sensing. However, the existing methods shows that the directionality of SPPs is always controlled by orthogonal polarizations. In fact, Spin-split of non-orthogonal polarizations is considerably necessary in some applications (e.g., polarization sensing and polarization splitting).

In this letter, we propose a method to control the directionality of SPPs by switching the handedness of input arbitrarily elliptically polarized light. Impressively, it turns out that the device can split two quite adjacent polarization components with a high extinction ratio. Our method provides opportunities for polarization sensing, polarization splitting and polarization-controlled plasmonic devices.

## 2. Principle and Structure

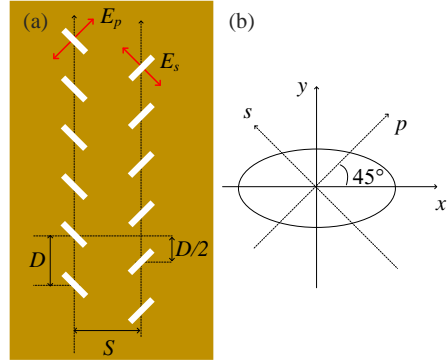

Fig. 1. (a) Schematic diagram and (b) input elliptically polarized light.

The directional SPPs coupler consists of two columns of air nanoslit arrays with 45° and 135° orientations respectively, as shown in Fig. 1(a). The two columns are placed at the symmetric positions of x-axis and the space between them is $S$. Here, we define the parameter of $\left|360°S/\lambda_{sp}-180°\right|$ as the elliptical angle of the device (EAD). The rectangular nanoslit arrays are fabricated in the 150-nanometer-thick gold surface and the dimensions of nanoslits are 50 nm * 200 nm. The period in y-direction labeled by $D$ is $\lambda_{SP}/2$. The input wavelength is 633 nm and the SPPs wavelength $\lambda_{SP}$ is calculated to be 606 nm [13]. Here, the incident polarization state is expressed as [13, 15]

$$\exp(-i\delta)\tan\Psi = \frac{E_s}{E_p}. \quad (1)$$

A nanoslit in a metal film selectively scatters incident light that is polarized perpendicular to it, giving rise to SPPs [13]. The SPPs can be expressed by

$$\begin{aligned}E_R &\propto E_p \exp(i2\pi S/\lambda_{sp}) + E_s \\ E_L &\propto -E_p - E_s \exp(i2\pi S/\lambda_{sp})\end{aligned}. \quad (2)$$

Where $E_R$ and $E_L$ indicate the SPPs fields propagating to the right and the left sides. Figure 2 illustrates the normalized SPPs power propagating to either side of the coupler as a function of the polarization state of the normally incident light according to Eq. (2) for three structures, i.e., EAD = 90°, 20°, 160° ($360°S/\lambda_{sp} = 90°, 160°, 340°$) respectively. The power is normalized to coupling power of circularly polarized light incident upon the device with structure parameter of $EAD$ =90°. It can be seen in Fig. 2 that the SPP excitations are different from each other. And interestingly, there are always two polarization states existing to make the SPP waves only propagate to left side and right side respectively, such as, Polarization B and Polarization E for EAD = 90°, Polarization C and Polarization D for EAD = 20°, Polarization A and Polarization F for EAD = 160°. It meets requirements of $\delta=\pm EAD$ and $\Psi = 45°$, which indicates the input light is a standard elliptically or circularly polarized light, as shown in Fig. 1(b). In the case, the Jones matrix can be rewritten as $[\sin\chi \quad \sigma i\cos\chi]^T$ in coordinate basis $[x \quad y]^T$ or $[1 \quad \exp(2\sigma i\chi)]^T$ in coordinate basis $[p \quad s]^T$. Here, $\delta=-2\sigma\chi$, $2\chi\in(0° \quad 180°)$, $\sigma$=1 and -1 represent the left and right handedness respectively. When $360°S/\lambda_{sp} = \pm 2\chi + 180°$, the SPPs field intensities can be written as

$$I_R \propto 4\sin^2[\chi(\sigma \mp 1)], \quad I_L \propto 4\sin^2[\chi(\sigma \pm 1)]. \tag{3}$$

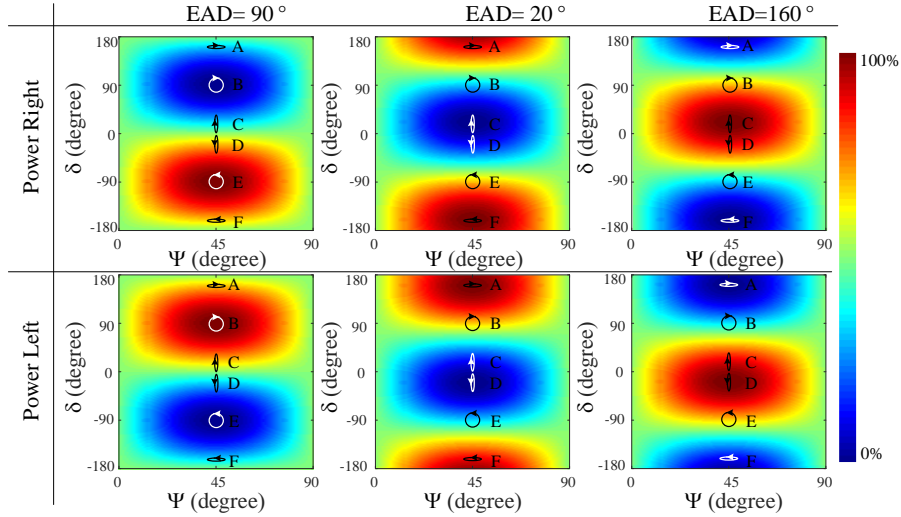

Fig. 2. Distribution of normalized power between the SPP waves propagating to either side of the coupler as a function of the polarization state of the normally incident light according to Eq. (1) and Eq. (2).

From Eq. (3), we can see that the spin-controlled directional coupling of SPPs is achieved and the coupling efficiency is proportional to $\sin^2(EAD)$. For arbitrarily elliptically polarized light, the spin-controlled directional coupling can be achieved equivalently by rotating the SPPs device by a certain angle, to make the fast axis or slow axis parallel to the columns of nanoslit arrays. For simplicity, we only consider the standard elliptically or circularly polarized light.

## 3. Results

We use a three-dimensional finite difference time domain (FDTD) method to simulate the SPPs field. Firstly, the influence of the space ($S$) on the SPPs field excited by each column of nanoslit arrays is studied under the periodic boundary condition. When a *p*-polarized or *s*-polarized unit light is incident upon the device, the power distributions and phase distributions at right and left sides (obtaining by two linear monitors placed at $x=\pm 5\mu m$) are shown in Fig.3 and Fig.4 (a). We can see that the powers from two polarizations are quite different when $S < 0.25\lambda_{sp}$, and become consistent with a power difference < 0.7 dB when $S \geq 0.25\lambda_{sp}$. The reason is that the two columns will have enhanced interaction with each other, such as mode coupling, when they are too close together. And the interaction can be ignored when they are apart. For the same reason, the phase distributions have the similar phenomena. From Fig.4 (a), we can see that the phase distributions have approximate slopes of $\pm 180°$. In order to clearly see the phase errors, we calculate the phase differences of two SPPs in the same side excited by two input polarizations. The phase errors are obtained by subtracting the phases caused by optical path differences (*S*) from the phase differences. More specifically, the right phase errors are expressed as $\phi_{R,p} - \phi_{R,s} - 360°S/\lambda$, where $\phi_{R,p}$ ($\phi_{R,s}$) is the monitored phase in the right monitor when the *p*-polarized (*s*-polarized) unit light is input. Similarly, the left phase errors are expressed as $\phi_{L,s} - \phi_{L,p} - 360°S/\lambda$. The phase errors in two sides are shown in Fig.4 (b). We can see that the phase errors are quite large when $S < 0.25\lambda_{sp}$ and become smaller than 5° when $S \geq 0.25\lambda_{sp}$. Based on the analysis, we should always choose $S \geq 0.25\lambda_{sp}$ in the following design. In the simulations, we chose $360°S/\lambda_{sp} = 180°-2\chi$ when $2\chi \leq 90°$ and $360°S/\lambda_{sp} = 180°+2\chi$ when $2\chi > 90°$ to avoid the undesired region.

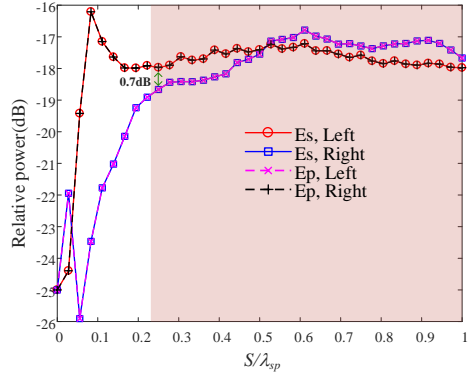

Fig. 3. Power distributions dependent on the space in left or right sides (labeled by "Left" or "Right") when a *p*-polarized (labeled by "Ep") or *s*-polarized (labeled by "Es") unit light is input.

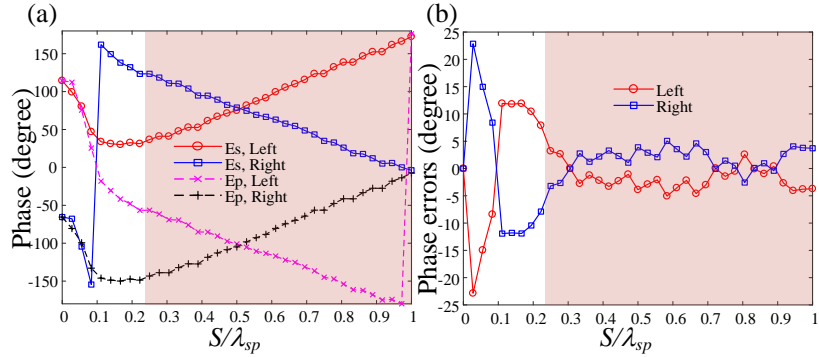

Fig. 4. (a) Phase distributions and (b) phase error distributions dependent on the space in left or right sides (labeled by "Left" or "Right") when a *p*-polarized (labeled by "Ep") or *s*-polarized (labeled by "Es") unit light is input.

We simulate the SPPs fields of various devices with 50 nanoslits in y-direction and a perfectly matched layer (PML) is used as the boundary treatment. Figure 5 presents the SPPs amplitude distributions for various spin-controlled directional coupling when the input parameter $2\chi$ =90°, 20°, 160°. From Figs. 5(a-c), one can see that our scheme can split various left-hand elliptically polarized (LEP) and right-hand elliptically polarized (REP) components to two opposite directions with a high polarization extinction ratio (PER), when the parameters of the device and the input light match with each other. Furthermore, the coupling efficiency is down obviously when the eccentricity of input elliptical polarization increases. It is reasonable because the coupling efficiency is proportional to. Besides, the coupling directions depending on handedness are inverse when $2\chi$ =20° and 160°, as shown in Figs. 5 (b) and (c). It is caused by the fact that the selection rule of structure parameter of EAD is changed, in other words, they choose the opposite signs in Eq. (3). In fact, the device in Fig. 5 (a) is a special case reported in Ref. [13], where the direction of SPPs was switched by left-hand and right-hand circularly polarized light (LCP and RCP). When the LCP or RCP light is incident upon the device with unmatched EAD, the effect of splitting becomes quite poor, as shown in Fig. 5 (d).

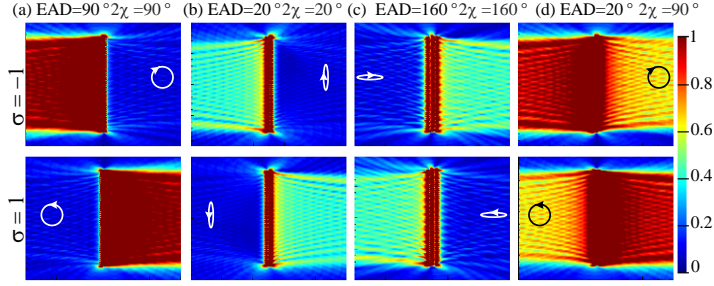

Fig. 5. The SPPs amplitude distributions for various spin-controlled directional coupling.

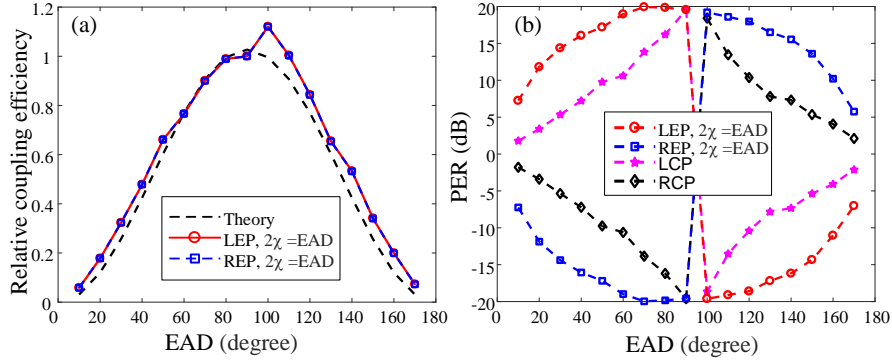

Fig. 6. (a) The relative coupling efficiency and (b) polarization extinction ratio for various spin-controlled directional coupling.

More detailed simulations for various perfect spin-controlled directional coupling devices are made to analyze the performances, such as the relative coupling efficiency and PER. Here we study the situation that the structure and the incident light match with each other (i.e., $2\chi$ =EAD). The results are depicted in Fig. 6. The coupling efficiency is obtained by sum of the power in two sides and normalized relative to the case of EAD=90°. From Fig. 6. (a), we can see that the coupling efficiency is approximately proportional to $\sin^2(EAD)$ and consistent to the theoretical results. In addition, the coupling efficiency reaches to the maximum value when EAD is slightly larger than 90°. The reason is that the power difference and phase difference when $S$ is near $0.75\lambda_{sp}$ is generally smaller than the ones when $S$ is near $0.25\lambda_{sp}$, which can be seen from Fig.3 and Fig. 4 (b). For more intuitive, we define PER as the power ratio between right side and left side. So the SPPs mainly propagate to the right side when PER>0 and to the left side when PER<0. The results are depicted in Fig. 6 (b). The device works very well and we can achieve a large value of PER when the EAD and the input polarization match (Labeled by "LEP" and "REP"), and still maintain good performance even if the eccentricity of input elliptical polarization is near 1. It implies that our method can split two quite adjacent polarization components to two opposite directions with a good PER. According to our results, the PER is about 7 dB when the EAD=10° (the eccentricity is > 0.996). The signs of PER are inverted when EAD < 90° and EAD > 90° since the selection rule of EAD is changed. Specifically, they choose the opposite signs in Eq. (3). The value of PER decreases obviously when the EAD doesn't match the input polarization, such as the results of Fig. 6 (b) (Labeled by "LCP" and "RCP") when the LCP or RCP light is incident upon various devices with unmatched EAD. Furthermore, the PER reaches its maximum value when EAD is near 90° and decreases as EAD is away from 90°. Also, the error tolerance of device parameters decreases when the eccentricity of input elliptical polarization increases. For example, when we only consider the phase error caused by the device, the phase error is quantized by $\Delta\phi$, then we have

$$|PER| = \sin^2[2\sigma\chi \pm \Delta\phi/2]/\sin^2(\Delta\phi/2) \qquad (4)$$

From Eq. (4), we can see that the $|PER|$ is proportional to $\sin^2[2\sigma\chi \pm \Delta\phi/2]$. And since the phase error is comparatively small as we showed before, it can be ignored when $2\chi$ is near $90^\circ$, which means that the input polarization is close to circular polarization. And the error tolerance will go bad when $2\chi$ is away from $90^\circ$, and the input polarization is close to linear polarization at the same time.

The coupling efficiency and the PER can be further improved by increasing the columns of air nanoslit arrays with a period of $\lambda_{sp}$ in x-direction. Figs. 7(a) and 7(b) show an example when EAD= 20° and the structure has five periods of air nanoslit arrays. The coupling efficiency is increased by a factor of about 21 and the PER is increased from 11.87 dB to 15 dB. The increased coupling efficiency is caused by the increased coupling region. The increasing PER is due to the truth that the SPPs excited by each column are constructively interfered in one direction. And in the other direction, the SPPs excited by each period are mainly caused by the power imbalance and phase errors. This part of SPPs will be not always constructively interfered.

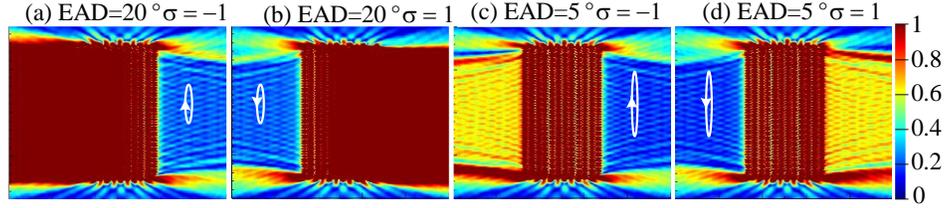

Fig. 7. The SPPs amplitude distributions for five periods of air nanoslit arrays.

Our method can control the directional coupling of SPPs by the handedness of arbitrarily elliptically polarized light. The coupling efficiency and the PER can be further improved by periodically increasing the columns of air nanoslit arrays. Our device can split two quite adjacent polarization components to two opposite directions exactly. Figs. 7(c) and 7(d) show an example when the EAD = 5° with five periods of arrays. We can see that the directional coupling of SPPs is still effective although the two elliptically polarized light are quite close to the same horizontal polarization. The calculated PER is about 7.8 dB. This property shows promising applications in special polarization sensing and special polarization splitting.

## 4. Conclusions

In conclusion, a method is presented to control the directionality of SPPs by switching the handedness of input arbitrarily elliptically polarized light. The coupling efficiency and the PER are studied in detail. We achieved a great value of PER when the EAD match the input polarization, and the PER still maintain great quality even if the eccentricity of input elliptical polarization is near 1. Consequently, the device can split two quite adjacent polarization components with a high extinction ratio. Our method provides opportunities for polarization sensing, polarization splitting and polarization-controlled plasmonic devices.

## References

[1] H. Kim, J. Park, S. W. Cho, S. Y. Lee, M. Kang, and B. Lee, "Synthesis and dynamic switching of surface plasmon vortices with plasmonic vortex lens," *Nano Lett,* vol. 10, pp. 529-36, Feb 10 2010.
[2] G. H. Yuan, Q. Wang, P. S. Tan, J. Lin, and X. C. Yuan, "A dynamic plasmonic manipulation technique assisted by phase modulation of an incident optical vortex beam," *Nanotechnology,* vol. 23, p. 385204, Sep 28 2012.
[3] W. Chen, D. C. Abeysinghe, R. L. Nelson, and Q. Zhan, "Experimental confirmation of miniature spiral plasmonic lens as a circular polarization analyzer," *Nano Lett,* vol. 10, pp. 2075-9, Jun 9 2010.
[4] W. Chen, D. C. Abeysinghe, R. L. Nelson, and Q. Zhan, "Plasmonic lens made of multiple concentric metallic rings under radially polarized illumination," *Nano Lett,* vol. 9, pp. 4320-5, Dec 2009.
[5] S. Yang, W. Chen, R. L. Nelson, and Q. Zhan, "Miniature circular polarization analyzer with spiral plasmonic lens," *Optics Letters,* vol. 34, pp. 3047-3049, 2009/10/15 2009.
[6] P. S. Tan, X. C. Yuan, J. Lin, Q. Wang, T. Mei, R. E. Burge*, et al.*, "Surface plasmon polaritons generated by optical vortex beams," *Applied Physics Letters,* vol. 92, p. 111108, Mar 17 2008.
[7] H. L. Zhou, J. J. Dong, Y. F. Zhou, J. H. Zhang, M. Liu, and X. L. Zhang, "Designing Appointed and Multiple Focuses With Plasmonic Vortex Lenses," *IEEE Photonics Journal,* vol. 7, pp. 1-7, Aug 2015.
[8] H. Zhou, J. Dong, J. Zhang, and X. Zhang, "Retrieving orbital angular momentum distribution of light with plasmonic vortex lens," *Scientific Reports,* vol. 6, p. 27265, 06/03/online 2016.
[9] G. Spektor, A. David, B. Gjonaj, G. Bartal, and M. Orenstein, "Metafocusing by a Metaspiral Plasmonic Lens," *Nano Lett,* vol. 15, pp. 5739-43, Sep 9 2015.
[10] H. Zhou, Y. Zhou, J. Zhang, and J. Dong, "Double-Slit and Square-Slit Interferences With Surface Plasmon Polaritons Modulated by Orbital Angular Momentum Beams," *IEEE Photonics Journal,* vol. 7, pp. 1-7, 2015.
[11] S. Xiao, F. Zhong, H. Liu, S. Zhu, and J. Li, "Flexible coherent control of plasmonic spin-Hall effect," *Nature Communications,* vol. 6, 2015.
[12] E.-Y. Song, S.-Y. Lee, J. Hong, K. Lee, Y. Lee, G.-Y. Lee*, et al.*, "A double-lined metasurface for plasmonic complex-field generation," *Laser & Photonics Reviews,* vol. 10, pp. 299-306, 2016.
[13] J. Lin, J. P. Mueller, Q. Wang, G. Yuan, N. Antoniou, X. C. Yuan*, et al.*, "Polarization-controlled tunable directional coupling of surface plasmon polaritons," *Science,* vol. 340, pp. 331-4, Apr 19 2013.
[14] L. L. Huang, X. Z. Chen, B. F. Bai, Q. F. Tan, G. F. Jin, T. Zentgraf*, et al.*, "Helicity dependent directional surface plasmon polariton excitation using a metasurface with interfacial phase discontinuity," *Light-Science & Applications,* vol. 2, p. e70, Mar 2013.
[15] Q. Xu, X. Zhang, Q. Yang, C. Tian, Y. Xu, J. Zhang*, et al.*, "Polarization-controlled asymmetric excitation of surface plasmons," *Optica,* vol. 4, p. 1044, 2017.
[16] S.-Y. Lee, K. Kim, S.-J. Kim, H. Park, K.-Y. Kim, and B. Lee, "Plasmonic meta-slit: shaping and controlling near-field focus," *Optica,* vol. 2, pp. 6-13, 2015/01/20 2015.
[17] O. You, B. Bai, L. Sun, B. Shen, and Z. Zhu, "Versatile and tunable surface plasmon polariton excitation over a broad bandwidth with a simple metaline by external polarization modulation," *Opt Express,* vol. 24, pp. 22061-73, Sep 19 2016.
[18] J. Chen, C. Sun, H. Li, and Q. Gong, "Ultra-broadband unidirectional launching of surface plasmon polaritons by a double-slit structure beyond the diffraction limit," *Nanoscale,* vol. 6, pp. 13487-93, Nov 21 2014.
[19] J. Yang, X. Xiao, C. Hu, W. Zhang, S. Zhou, and J. Zhang, "Broadband surface plasmon polariton directional coupling via asymmetric optical slot nanoantenna pair," *Nano Lett,* vol. 14, pp. 704-9, Feb 12 2014.
[20] B. Zhu, G. Ren, Y. Gao, B. Wu, C. Wan, and S. Jian, "Graphene circular polarization analyzer based on unidirectional excitation of plasmons," *Optics Express,* vol. 23, pp. 32420-32428, 2015/12/14 2015.
[21] J. P. Balthasar Mueller, K. Leosson, and F. Capasso, "Ultracompact metasurface in-line polarimeter," *Optica,* vol. 3, pp. 42-47, 2016/01/20 2016.
[22] J. P. Mueller, K. Leosson, and F. Capasso, "Polarization-selective coupling to long-range surface plasmon polariton waveguides," *Nano Lett,* vol. 14, pp. 5524-7, Oct 8 2014.
[23] S. Y. Lee, K. Kim, G. Y. Lee, and B. Lee, "Polarization-multiplexed plasmonic phase generation with distributed nanoslits," *Opt Express,* vol. 23, pp. 15598-607, Jun 15 2015.
[24] F. Ye, M. J. Burns, and M. J. Naughton, "Symmetry-Broken Metamaterial Absorbers as Reflectionless Directional Couplers for Surface Plasmon Polaritons in the Visible Range," *Advanced Optical Materials,* vol. 2, pp. 957-965, 2014.
[25] Q. Xu, X. Zhang, Y. Xu, C. Ouyang, Z. Tian, J. Gu*, et al.*, "Polarization-controlled surface plasmon holography," *Laser & Photonics Reviews,* vol. 11, p. 1600212, 2017.
[26] Y. Lefier and T. Grosjean, "Unidirectional sub-diffraction waveguiding based on optical spin-orbit coupling in subwavelength plasmonic waveguides," *Opt Lett,* vol. 40, pp. 2890-3, Jun 15 2015.
[27] S.-Y. Lee, H. Yun, Yohan, and B. Lee, "Switchable surface plasmon dichroic splitter modulated by optical polarization," *Laser & Photonics Reviews,* vol. 8, pp. 777-784, 2014.
[28] A. Pors, M. G. Nielsen, T. Bernardin, J.-C. Weeber, and S. I. Bozhevolnyi, "Efficient unidirectional polarization-controlled excitation of surface plasmon polaritons," *Light Sci Appl,* vol. 3, p. e197, 08/15/online 2014.